\documentclass[%
]{revtex4-2}

\usepackage{rotating}
\usepackage{lipsum}
\usepackage{graphicx}
\usepackage{dcolumn}
\usepackage{bm}
\usepackage{color}
\usepackage{diagbox}
\usepackage[table]{xcolor}
\usepackage{tabularx}
\usepackage{makecell}
\usepackage[none]{hyphenat}
\usepackage{lineno}
\usepackage{mathrsfs}

\definecolor{Navy}{RGB}{54,100,139}
\usepackage[colorlinks,
linkcolor=Navy,
anchorcolor=Navy,
citecolor=Navy]{hyperref}
\usepackage[version=4]{mhchem}
\usepackage{amsmath}
\usepackage{bm}
\usepackage{placeins}
\usepackage{changes}
\definechangesauthor[name={HI}, color=orange]{HI}

\begin{document}


\title{
       Tunable topological phases 
       in nanographene-based spin-½ alternating-exchange Heisenberg chains  \\
}

\author{Chenxiao Zhao$^{1,}$\footnote{\label{note1}These authors contributed equally.}, Gon\c{c}alo Catarina$^{1,2,}$\footref{note1},  Jin-Jiang Zhang$^{3,5,}$\footref{note1}, Jo\~{a}o C. G. Henriques$^{2,4}$, Lin Yang$^{3,5}$, Ji Ma$^{3,5}$, Xinliang Feng$^{3,5,}$\footnote{\label{note2}Corresponding authors: xinliang.feng@tu-dresden.de, pascal.ruffieux@empa.ch, joaquin.fernandez-rossier@inl.int}, Oliver Gröning$^{1}$, Pascal Ruffieux$^{1,}$\footref{note2},  Joaqu\'{i}n Fern\'andez-Rossier$^{2,}$\footref{note2}$^,$\footnote{On permanent leave from Departamento de F\'{i}sica Aplicada, Universidad de Alicante, San Vicente del Raspeig, Spain.}, Roman Fasel$^{1,6}$}
\affiliation{$^{1}$Empa -- Swiss Federal Laboratories for Materials Science and Technology, D\"{u}bendorf, Switzerland.}
\affiliation{$^{2}$International Iberian Nanotechnology Laboratory, Braga, Portugal.}
\affiliation{$^{3}$Faculty of Chemistry and Food Chemistry, and Center for Advancing Electronics Dresden, Technical University of Dresden, Dresden, Germany.}
\affiliation{$^{4}$Universidade de Santiago de Compostela, Santiago de Compostela, Spain.}
\affiliation{$^{5}$Max Planck Institute of Microstructure Physics, Halle, Germany.}
\affiliation{$^{6}$University of Bern, Bern, Switzerland.}
%
%

\begin{abstract}
	\textbf{Unlocking the potential of topological order within many-body spin systems has long been a central pursuit in the realm of quantum materials. Despite extensive efforts, the quest for a versatile platform enabling site-selective spin manipulation, essential for tuning and probing diverse topological phases, has persisted. Here, we utilize on-surface synthesis to construct spin-½ alternating-exchange Heisenberg (AH) chains\textsuperscript{\cite{duffy1968theory}} with antiferromagnetic couplings $J_1$ and $J_2$ by covalently linking Clar’s goblets---nanographenes each hosting two antiferromagnetically-coupled unpaired electrons\textsuperscript{\cite{mishra2020topological}}. Utilizing scanning tunneling microscopy, we exert atomic-scale control over the spin chain lengths, parities and exchange-coupling terminations, and probe their magnetic response by means of inelastic tunneling spectroscopy. 
    Our investigation confirms the gapped nature of bulk excitations in the chains, known as triplons\textsuperscript{\cite{triplon}}. Besides, the triplon dispersion relation is successfully extracted from the spatial variation of tunneling spectral amplitudes. 
    Furthermore, depending on the parity and termination of chains, we observe varying numbers of in-gap $S=$~½ edge spins, enabling the determination of the degeneracy of distinct topological ground states in the thermodynamic limit—either 1, 2, or 4. By monitoring interactions between these edge spins, we identify the exponential decay of spin correlations. Our experimental findings, corroborated by theoretical calculations, present a phase-controlled many-body platform, opening promising avenues toward the development of spin-based quantum devices. }

\end{abstract}
\maketitle


\sloppy{}

Strongly correlated quantum many-body systems and topological phases stand as two pivotal focal points in modern condensed matter physics and have been intensively explored in quantum dots\textsuperscript{\cite{kiczynski2022engineering}}, cold atoms\textsuperscript{\cite{sompet2022realizing}}, molecules\textsuperscript{\cite{mishra2021observation}}, twisted bilayer structures\textsuperscript{\cite{xie2021fractional}}, and conventional crystals with exotic phases, such as Kitaev materials\textsuperscript{\cite{trebst2022kitaev}} and Wigner crystals\textsuperscript{\cite{li2021imaging}}. Their interplay gives rise to novel phases like the topological Kondo insulator\textsuperscript{\cite{topologicalkondo}}, topological superconductivity\textsuperscript{\cite{hasan2010colloquium, qi2011topological}}, and
topological spin liquids\textsuperscript{\cite{QSL,QSL2,grohol2005spin,sompet2022realizing}}. 
A paradigmatic example of a topological spin system is the spinful many-body analog of the Su–Schrieffer–Heeger (SSH) model\textsuperscript{\cite{su1979solitons}}, referred to as the one-dimensional (1D) spin-½ alternating-exchange Heisenberg (AH) model\textsuperscript{\cite{hida1992crossover}}: 
\begin{equation}
	\begin{aligned}
		\hat{\mathcal{H}}= \sum\limits_{i} (J_1\hat{\bm{S}}_{2i-1}\cdot\hat{\bm{S}}_{2i} +J_2\hat{\bm{S}}_{2i}\cdot\hat{\bm{S}}_{2i+1})
		\label{eq1}
	\end{aligned}
\end{equation}
where $J_1>0$ and $J_2>0$ denote the two alternating antiferromagnetic couplings, and $\hat{\bm{S}}_i$ denotes the vector of spin-½ operators at site $i$.  Throughout the manuscript we use the convention $J_2>J_1$ without loss of generality. Given this condition, the model belongs to the same topological phase as the spin-1 Haldane chains, featuring gapped excitations in the bulk and non-local string order parameters\textsuperscript{\cite{hida1992crossover}}. For AH chains with open boundary conditions (OBC) terminated by the weaker coupling $J_1$, the system has in-gap edge excitations that, in the thermodynamic limit, become gapless, leading to a fourfold degenerate ground state (Fig. \ref{fig1}\textcolor{Navy}{a}). In contrast,  and at odds with open-ended spin-1 Haldane chains, AH chains terminated by the stronger coupling $J_2$ have a non-degenerate ground state without edge spins, and AH chains with mixed terminations have a twofold degenerate ground state, with a dangling spin localized at the $J_1$ terminus (Fig. \ref{fig1}\textcolor{Navy}{a}). Therefore, depending on chain parity (i.e., odd or even number of spin sites) and termination, the degeneracy of the ground state can be 1, 2, or 4, reflecting the underlying symmetry-protected topological order\textsuperscript{\cite{chen11}}. In the bulk, all three phases are characterized by a valence bond (VB) crystal ground state\textsuperscript{\cite{VBSorder}}, with the VB pairs pinned by the stronger $J_2$ coupling. The elemental bulk excitation is a bosonic spin-1 quasiparticle, called triplon\textsuperscript{\cite{balents2010spin,drost2023real}}. 

Over the past decades, extensive experimental efforts have been dedicated to 
realize this spin-½ AH model, primarily concentrating on quasi-1D spin arrays embedded in three-dimensional (3D) crystals or polymers\textsuperscript{\cite{diederix1979theoretical,bonner1983alternating,garrett1997magnetic,forsyth1991antiferromagnetic,lake1997dimer,waki2004observation,valentine1974interdimer,bray1975observation,jacobs1976spin,castilla1995quantum,riera1995magnetic,hase1993observation}}. However, spin-phonon coupling and residual exchange interactions between spin arrays are inevitable in these 3D crystals. 
Furthermore, the absence of practical methods to control the chain length, parity, and termination type within 3D crystals hinders systematic investigations for the spin-½ AH model, particularly for their topological edge effects.
Recently, the fabrication of artificial many-body SSH chains with quantum dots has been reported\textsuperscript{\cite{kiczynski2022engineering}}, but experiments were carried out at quarter-filling, whereas the mapping to the AH model requires half-filling.
So far, an ideal platform for realizing the spin-½ AH chain has yet to be established, let alone a controllable quantum phase transition in such a many-body system.

Recent developments in on-surface synthesis\textsuperscript{\cite{cai2010atomically,ruffieux2016surface,rizzo2018topological,blackwell2021spin}} and the ensuing realization of magnetic nanographenes have shown great potential of realizing quantum spin systems\textsuperscript{\cite{ortiz2019exchange,mishra2020topological,mishra2021large,cheng2022surface,wang2022aza,mishra2021observation}}. 
In this study, we harness the potential of on-surface synthesis to covalently connect goblet-shaped magnetic nanographenes\textsuperscript{\cite{mishra2020topological}}  into chains on a Au(111) surface (see \textcolor{Navy}{Methods} and Extended Data Fig. \ref{FigE7} for details of sample preparation), which we show to be a paradigmatic realization of the spin-½ AH model. Scanning tunneling microscopy (STM) enables spatial- and energy-resolved characterization of spin excitations by inelastic electron tunneling spectroscopy\textsuperscript{\cite{madhavan1998tunneling,hirjibehedin2006spin,mishra2020topological}}, and also provides precise spin manipulation enabling the control of chain length, parity, and termination\textsuperscript{\cite{wang2022aza,Tip-induced}}. This allows us to explore three distinct phases, distinguished by their ground state degeneracies, that become apparent by the presence or lack of edge excitations. Moreover, the dispersive nature of triplons is successfully extracted, for the first time in an STM experiment, by leveraging quantum confinement effects in finite-length AH chains.

\section{Realizing the AH model}

\begin{figure*}
	\includegraphics[width=16 cm]{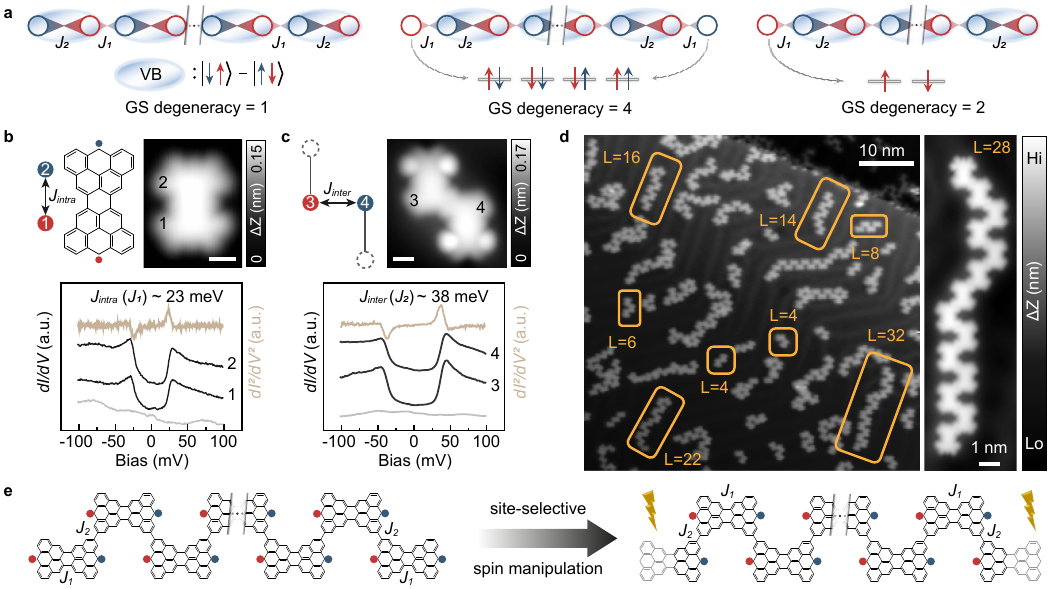}
	\caption{\label{fig1}\textbf{Realization of AH model in goblet chains.}
    \textbf{a},	Schematic illustrations of AH chains with different terminations. The corresponding ground state degeneracy is displayed below. VB pairs are represented by ellipses. 
    \textbf{b} and \textbf{c}, Schematic structures and STM images of a goblet ($V_{bias}=-0.05$ V, $I_{set}=500$ pA) and a modified goblet dimer ($V_{bias}=-43$ mV, $I_{set}=500$ pA), featuring exchange couplings $J_{intra}$ and $J_{inter}$, respectively. The differential conductance spectra ($dI/dV$) taken on different spin sites are shown in the lower panel ($I_{set}=500$ pA), with the background spectra on the Au(111) substrate shown by gray curves. The averaged $d^2I/dV^2$ spectra are shown by light brown curves. White scale bars denote 0.5 nm. $dI/dV$ spectra are taken with root mean squared modulation voltage $V_{rms}=2$ mV.
    \textbf{d}, Large-scale topographic image of goblet chains synthesized on Au(111) ($V_{bias}=-0.10$ V, $I_{set}=50$ pA). Chains with different lengths $L$ (number of spin-½ sites) are marked out. A typical long chain is shown in the right panel ($V_{bias}=-0.30$ V, $I_{set}=500$ pA). 
    \textbf{e}, Structural illustration of tip-controlled spin manipulation, switching chain termination from $J_1$ to $J_2$, where $J_1=J_{intra}$ and $J_2=J_{inter}$.
	 }
\end{figure*}

As a first step, we validate the realization of the AH model in our nanographene system. The fundamental building block is a Clar's goblet (C$_{38}$H$_{18}$, hereafter referred to as goblet), which harbors two ½-spins that are antiferromagnetically coupled to each other\textsuperscript{\cite{mishra2020topological,ortiz2019exchange}}, as illustrated in Fig. \ref{fig1}\textcolor{Navy}{b}. Differential conductance ($dI/dV$) spectra taken on both sides of the goblet exhibit two steps that are symmetric with respect to zero bias (Fig. \ref{fig1}\textcolor{Navy}{b}), representing the excitation from the singlet ground state to the triplet excited state\textsuperscript{\cite{mishra2020topological,ortiz2019exchange}}. The corresponding step energy, indicated by peak positions in the second derivative $d^2I/dV^2$, reveals the magnitude of the exchange coupling $J_{intra}$ to be $\sim$23 meV, in line with previous reports\textsuperscript{\cite{mishra2020topological}}. 
Connecting two goblets into a dimer gives rise to the nearest ($J_{inter}$), next-nearest ($J_{nn}$), and third-nearest ($J_{nnn}$) neighbor couplings (Extended Data Fig. \ref{FigE1}). To isolate and determine these exchange couplings, hydrogenation and tip-induced dehydrogenation techniques are applied to the dimer\textsuperscript{\cite{mishra2020topological,Tip-induced}}, by which spin sites are selectively switched off (see \textcolor{Navy}{Methods}). 
A modified dimer involving only $J_{inter}$ reveals its antiferromagnetic nature with a coupling strength of $\sim$38 meV (Fig. \ref{fig1}\textcolor{Navy}{c}), where the symmetric excitation steps around zero bias correspond to singlet-triplet excitations. 
Further investigation into dimers involving $J_{nn}$ and $J_{nnn}$ discloses their negligible magnitude (Extended Data Fig. \ref{FigE1}). 
We have also verified that $J_{inter}$ is not sensitive to the relative configuration of neighboring molecules, which can be either inversion-symmetric (\textit{trans}, Fig. \ref{fig1}\textcolor{Navy}{c}) or mirror-symmetric (\textit{cis}, Extended Data Fig. \ref{FigE1}). 

Both the magnitude and the sign of the measured exchange interactions are well-reproduced by a fermionic Hubbard model solved within the configuration interaction framework with a complete active space (CAS) approximation\textsuperscript{\cite{ortiz2019exchange,krane2023exchange}} (Extended Data Fig. \ref{FigE2}, see \textcolor{Navy}{Methods} for details). 
Therefore, we consider only two alternating nearest-neighbor exchange couplings $J_1=J_{intra}$ and $J_2=J_{inter}$ in goblet chains. 
Given that the low-energy states, isomorphic to those of the pure spin Hamiltonian, are well separated from higher energy states involving charge fluctuations, and that the magnetic anisotropy in nanographene-based spin chains is negligible\textsuperscript{\cite{lado2014magnetic,mishra2021observation}}, we thus anticipate that the low-energy physics of the goblet chain can be described by the spin-½ AH model as presented in equation (\ref{eq1}).

Fig. \ref{fig1}\textcolor{Navy}{d} shows an overall topographic image of goblet chains synthesized on the Au(111) surface. Chains so obtained directly are terminated by $J_1$, associated with
complete goblets at the ends. $J_2$-terminated chains can be realized by passivating the outer-most spin
sites using hydrogenative or dehydrogenative passivation\textsuperscript{\cite{Tip-induced}} (Fig. \ref{fig1}\textcolor{Navy}{e}, see \textcolor{Navy}{Methods} for details). This permits us to undertake a systematic study of spin chains with different types of terminations.

\section{Even- and odd-Haldane phases}
\begin{figure*}[t]
	\includegraphics[width=16cm]{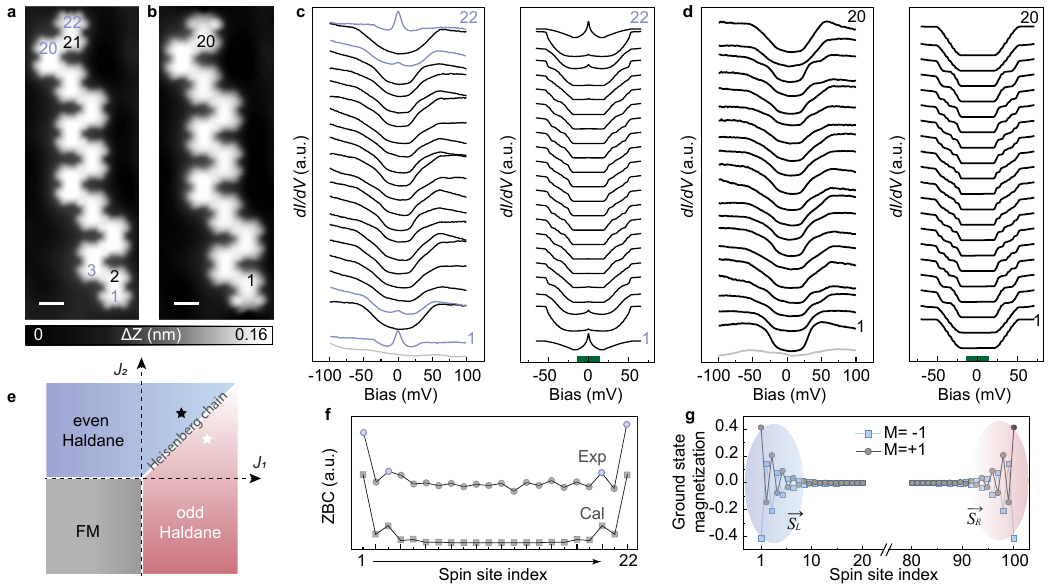}
	\caption{\label{Fig2}\textbf{Characterization of $J_1$- and $J_2$-terminated chains.}
		\textbf{a}, Topographic image of a $J_1$-terminated goblet chain with $L=22$. $V_{bias}=-58$ mV, $I_{set}=500$ pA. \textbf{b}, Topographic image of a $J_2$-terminated chain with $L=20$,  obtained by passivating the two edge spins of the chain shown in \textbf{a}. $V_{bias}=-100$ mV, $I_{set}=500$ pA.  White scale bars denote 2 nm. \textbf{c} and \textbf{d}, $dI/dV$ spectra ($I_{set}=500$ pA, $V_{rms}=2$ mV) taken at each spin site of the chains shown in \textbf{a} and \textbf{b}, respectively, with the gray spectra taken on the Au(111) substrate as a reference. The simulated spectra are shown on the right sides. The green bars indicate the energy range where third-order scattering processes are taken into account. 
        \textbf{e}, Phase diagram for the  AH model as a function of $J_1$ and $J_2$, assuming $J_1$ terminations. In experiments, goblet chains terminated by $J_1$ ($J_2$) reside in the even- (odd-) Haldane phase, as indicated by the black (white) star. 
        \textbf{f} Spatially-resolved ZBC extracted from both experimental and calculated $dI/dV$ spectra in \textbf{c}.  
        \textbf{g}, DMRG calculation of the local magnetic moments for the $M = \pm 1$ magnetization sectors of the fourfold degenerate ground state of a $J_1$-terminated goblet chain with $L=100$. The exponentially-localized emergent left ($\vec{{S}}_L$) and right ($\vec{{S}}_R$) edge spins are marked out.}
\end{figure*}

Fig. \ref{Fig2}\textcolor{Navy}{a} shows an $L=22$  $J_1$-terminated goblet chain (where $L$ denotes the number of spin-½ sites in the chain). The corresponding $dI/dV$ spectra taken on each spin site are shown in the left panel of Fig. \ref{Fig2}\textcolor{Navy}{c}. A striking feature is the presence of zero-bias peaks (ZBPs) at both chain termini (sites 1 and 22), together with a nearly constant gap in the bulk of the chain.  A faint ZBP is also detected at the third site from the edge (sites 3 and 20), which is notably evident in the spatially-resolved zero-bias conductance (ZBC) extracted from the $dI/dV$ spectra (Fig. \ref{Fig2}\textcolor{Navy}{f}), indicating an oscillating decay of the ZBP. 
The $dI/dV$ spectra are modeled using the eigenstates of the spin chain, obtained by exact diagonalization (ED) of the Hamiltonian (\ref{eq1}). 
We treat tip-surface tunneling up to second order\textsuperscript{\cite{Rossier2009ITS}}, supplemented by a third-order correction\textsuperscript{\cite{ternes2015spin}} near zero bias (within $\pm$15 mV), to capture both spin excitations and zero-bias Kondo peaks (see \textcolor{Navy}{Methods} for details). 
As shown in the right panel of Fig. \ref{Fig2}\textcolor{Navy}{c}, the calculated spectra reproduce
the main experimental results, namely the bulk gap, the presence of multiple excitations above the gap,  the ZBPs at the edges and, remarkably, their oscillating decay into the bulk (Fig. \ref{Fig2}\textcolor{Navy}{f}). The agreement between experiments and the simulation validates the use of the AH model to describe our spin chains and unveils the Kondo-screened nature of the ZBPs stemming from the edge spins.  
The presence of two edge spins in the $L=22$ chain leads to a nearly fourfold-degenerate ground state (singlet-triplet splitting $<5$ $\mu$eV, see Extended Data Fig. \ref{FigE3}), reflecting the weak interaction between them. This is in line with the exponential localization of the edge spins, as evidenced by the ground state magnetization of a long chain calculated using the density-matrix renormalization group (DMRG) method\textsuperscript{\cite{white1992density}} (Fig. \ref{Fig2}\textcolor{Navy}{g}).

Switching off the outer-most spin sites of the chain converts the terminations to $J_2$. As shown in Fig.  \ref{Fig2}\textcolor{Navy}{b}, the original terminal spin sites are passivated by tip-induced dehydrogenation, which now show the typical ``big eyes'' feature of dehydrogenative passivation\textsuperscript{\cite{Tip-induced}}. As a consequence, the effective length of the chain reduces to 20 spin sites. The corresponding $dI/dV$ spectra taken along this modified chain are presented in Fig. \ref{Fig2}\textcolor{Navy}{d}, where ZBPs are absent. Instead, a significant gap with a broad slope is observed at both terminations (1 and 20), which is explained by the dense multi-step features revealed by the simulated spectra (right panel of Fig.  \ref{Fig2}\textcolor{Navy}{d}).  Individual steps can be resolved from the third site onwards (between sites 3 and 18). The bulk gap retains a similar value to that observed before passivation. Notably, without the influence of edge spins, successive spectra along the chain display a consistent pattern indicative of dimerization, meaning that spin sites connected through $J_2$ share identical properties. This spectroscopic manifestation of dimerization is also evident in the bulk of the $J_1$-terminated chain. The presence of dimerization and the bulk gap unveil the VB crystal nature in the bulk of the system, where the VB pairs are spatially confined by the $J_2$ coupling (see Fig. \ref{fig1}\textcolor{Navy}{a} and \textcolor{Navy}{Supplementary Note 1}). 
 
For infinitely long chains, i.e., in the thermodynamic limit, the $J_1$-terminated chain has a fourfold degenerate ground state and can be adiabatically deformed into the spin-1 Haldane chain in the limit $J_1 \rightarrow -\infty$ \textsuperscript{\cite{hida1992crossover}}. In contrast, the $J_2$-terminated chain maps into a spin-1 Haldane chain passivated with two ½ side spins that gap out the edge states.
As a result, distinct order parameters are needed to characterize the $J_1$- and $J_2$-terminated chains. Here we adopted the even- and odd-Haldane string order parameters\textsuperscript{\cite{wang2013topological}}, and confirmed that the $J_1$- and $J_2$-terminated goblet chains reside at the even- and odd-Haldane phase, respectively (Fig. \ref{Fig2}\textcolor{Navy}{e}, details in \textcolor{Navy}{Supplementary Note 2}). The phase boundary between them is at $J_1=J_2$, where the system becomes a gapless uniform Heisenberg chain.

\section{Quantum confinement of dispersive triplons}
\begin{figure*}[t]
	\includegraphics[width=16 cm]{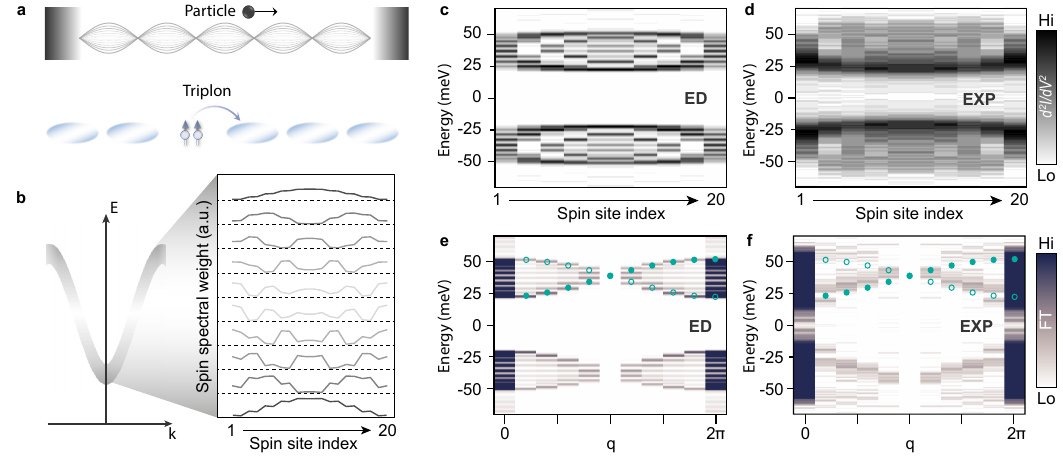}
	\caption{\label{Fig3}\textbf{Dispersive triplon in the AH chain.}
	\textbf{a}, Schematic illustration depicting the analogy between a confined standing wave of a particle in a box and a triplon excitation in a finite VB chain with OBC.
 \textbf{b}, Sketch of the dispersion relation of the one-triplon band in the AH chain, together with the calculated spatially-resolved spin spectral weights for the 10 one-triplon modes of an $L=20$ odd-Haldane chain, with energy increasing from bottom to top. 
 \textbf{c}, Calculated $d^2I/dV^2$ spectra of an $L=20$ odd-Haldane chain, obtained by differentiation of the spectra shown in the right panel of Fig. \ref{Fig2}\textcolor{Navy}{d}. \textbf{d}, Spatial- and energy-symmetrized $d^2I/dV^2$ spectra for the $L=20$ odd-Haldane chain shown in Fig. \ref{Fig2}\textcolor{Navy}{b}. Spectra are averaged within each VB pair, in light of the exclusive existence of a triplon within the VB (see \textcolor{Navy}{Supplementary Note 1}). \textbf{e} and \textbf{f}, Discrete FT of \textbf{c} and \textbf{d}, respectively.  The one-triplon dispersion relation, obtained with the triplet-wave expansion method and adapted to the case of OBC, is represented by the filled green circles, with the folded part denoted by the empty circles.	
	}
\end{figure*}

Although distinct in terms of the emergence of edge spins, AH chains in both even- and odd-Haldane phases share a VB crystal ground state in the bulk, with elemental excitations that are triplon-quasiparticles carrying a local spin-1 and hopping between VB pairs (Fig. \ref{Fig3}\textcolor{Navy}{a}). Here we focus on the low-lying one-triplon modes (one triplon within the chain), which play a dominant role in single-site spin-flip excitation experiments, such as in STM-based measurements (see Extended Data Fig. \ref{FigE4}). In the thermodynamic limit, the one-triplon band shows a cosine-like dispersion\textsuperscript{\cite{triplon}}: 
\begin{equation}
	\frac{E(k)}{J_2} = 1-\frac{\alpha}{2} \text{cos} (k) - \frac{\alpha^2}{8} \text{cos} (k)[2+\text{cos} (k)] 
 \label{eq3}
\end{equation}
where $\alpha$ denotes the $J_1/J_2$ ratio. 
In a finite chain with OBC, analogously to the particle-in-a-box model (Fig. \ref{Fig3}\textcolor{Navy}{a}), confinement leads to the development of standing triplon waves with quantized momentum given by $k = \frac{2\pi n}{D}$ ($n=1,2,...,D$), where $D$ denotes the number of VB pairs. Moreover, the OBC induces a $k$-space folding, given by $E_{OBC}(k) = E(k/2)$, as confirmed by the results of linear spin-wave theory\textsuperscript{\cite{ochsenbein2007standing}}. 
The resulting spatial and energy modulations of the triplon modes are reflected in spatially-resolved spin spectral weights\textsuperscript{\cite{Rossier2009ITS}}. For reference, the calculated spin spectral weights for an $L=20$ AH chain in the odd-Haldane phase are shown in Fig. \ref{Fig3}\textcolor{Navy}{b}, revealing the sinusoidal character of the $10$ one-triplon modes. 
Representing the amplitude of these spin spectral weights in a energy-site plot gives the dynamic spin structure factor (Fig. \ref{Fig3}\textcolor{Navy}{c}), which is proportional to the spatially-resolved $d^2I/dV^2$ spectra in the experiment\textsuperscript{\cite{drost2023real}}, as shown in Fig. \ref{Fig3}\textcolor{Navy}{d}. 
Both positive and negative energy branches are obtained, reflecting the symmetric spin excitations in STM measurements. Despite significant experimental broadening, the $d^2I/dV^2$ spectra obtained in the experiment match well with the calculation. A notable feature is the approximate symmetry of the positive and negative branches with respect to $J_2$ and $-J_2$, respectively, corresponding to the energy of the localized triplon in the fully dimerized case ($J_1=0$).
Introducing a non-zero $J_1$ imparts a dispersive character to these triplons, leading to the renormalization and splitting of their energies\textsuperscript{\cite{drost2023real}} (see Extended Data Fig. \ref{FigE4}). The asymmetry in intensity between the lower ($|E|<J_2=38$ meV) and higher ($|E|> J_2=38$ meV) energy windows in experiments is attributed to the increasing contribution of elastic tunneling conductance with increasing bias\textsuperscript{\cite{d2I/dV2}}. 

To filter out the non-modulated noise and extract dispersive information of triplons, a discrete Fourier transform (FT) was applied to the $d^2I/dV^2$ spectra. As illustrated in Figs. \ref{Fig3}\textcolor{Navy}{e} and \textcolor{Navy}{f}, both the calculated and the experimental results exhibit a cosine-like pattern, qualitatively aligning with the dispersion relation $E_{OBC}(q)$ ($q$ denotes the wave vector in the folded $k$-space), superimposed as solid green circles.
It is noteworthy that, aside from the $k$-space folding, a $q$-space folding emerges due to the discrete FT, introducing a replica wave (empty green circles) characterized by a $\pi$-shift of $E_{OBC}(q)$. 

\section{Spin correlation and manipulation}
\begin{figure*}
	\includegraphics[width=16 cm]{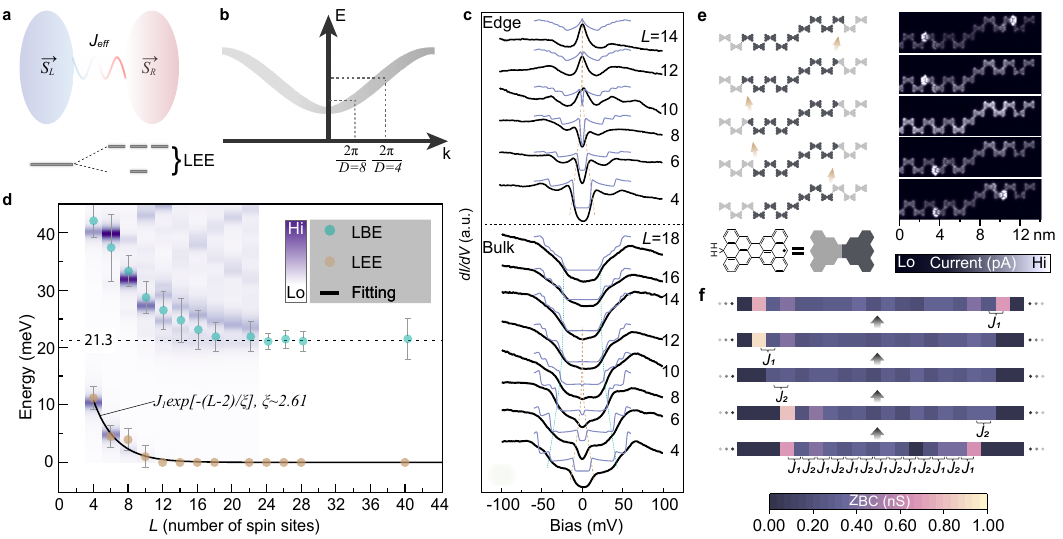}
	\caption{\label{Fig4}\textbf{Finite size effects and manipulation of the AH chain.}
		\textbf{a}, Schematic illustration of the effective coupling between edge spins in finite even-Haldane chains, resulting in the splitting of the fourfold degenerate ground state into singlet and triplet states,  identified as the lowest-energy edge excitation (LEE). \textbf{b}, Scheme depicting the minimum momentum for chains of varying lengths, leading to the length dependence of the lowest-energy bulk excitation (LBE). \textbf{c}, $dI/dV$ spectra taken at the edges and in the bulk of even-Haldane chains with different lengths  (see Extended Data Fig. \ref{FigE5} for details). Simulated spectra, up to third-order in scattering processes, are represented by the purple curves. Guidelines for the variation trend of LEE and LBE are shown by green and brown dashed curves, respectively. \textbf{d}, Evolution of the LEE and the LBE with chain length (see Extended Data Fig. \ref{FigE5} for original $dI/dV$ spectra). The calculated excitations on the bulk site are represented by the purple color map, where the color indicates the spin spectral weight. The LEEs can be captured by an exponential decay, as shown by the black solid line. Theoretical values of the bulk gap, obtained through both triplet-wave expansion and multiprecision methods, are indicated by the gray dashed line at 21.3 meV. Error bars for LEEs represent the measurement error in estimating the peaks of $d^2I/dV^2$ spectra, while error bars for LBEs also encompass the effect from the multi-excitation steps nearby.    \textbf{e}, Schematic drawing of the sequential activation process for a pre-hydrogenated goblet chain, where the activated spin sites are denoted by dark gray color and the passivated spin sites are denoted by light gray color. The corresponding bond-resolved current images are shown aside ($V_{bias}=-5$ mV). 
        \textbf{f}, ZBC extracted from the $dI/dV$ spectra taken for the activated segments during the sequential activation shown in \textbf{e}  (see Extended Data Fig. \ref{FigE6} for detailed $dI/dV$ spectra).
	}
\end{figure*}

For AH chains whose length is shorter or comparable to the spin correlation length ($\xi$), finite size effects and short-range interactions cannot be disregarded. A direct consequence in the even-Haldane phase is the effective exchange interaction  ${J}_{\rm eff}(L) \vec{{S}}_L\cdot\vec{{S}}_R$ between the emergent edge spins (Fig. \ref{Fig4}\textcolor{Navy}{a}). As a result, the fourfold degenerate ground state in the thermodynamic limit ($L \gg \xi$) splits into singlet and triplets states\textsuperscript{\cite{kennedy1990exact}}, with the splitting detectable as the lowest-energy edge excitation (LEE). The $dI/dV$ edge and bulk spectra of chains with different lengths are shown in Fig. \ref{Fig4}\textcolor{Navy}{c}. For chains with $L=4, 6, 8$, the LEE can be clearly resolved as a gap in the edge spectra. For the $L=10$ chain, the low-energy feature becomes strongly renormalized and, for the $L\geq12$ chains, the gap is replaced by a ZBP, reflecting that Kondo exchange between the emergent edge spins and the substrate dominates in the competition with the inter-edge exchange. 
As shown in Fig. \ref{Fig4}\textcolor{Navy}{d}, the extracted LEE shows an exponential decay with $L$.
This can be understood from the exponential decay of the inter-edge coupling, ${ J}_{\rm eff}(L) = J_1 \mathrm{e}^{-(L-2)/\xi}$, where $\xi = 2.61$ is the correlation length obtained from the fitting to the ED calculation. The limited correlation length establishes that, for a chain with mixed terminations, the edge with the $J_2$ termination shows even-Haldane phase character, namely the presence of an edge spin and its oscillating decay toward the bulk, whereas the edge with the $J_1$ termination has odd-Haldane phase character, provided that $L\gg\xi$ (see Extended Data Fig. \ref{FigE6}).  

Moreover, the lowest-energy bulk excitation (LBE) also exhibits a decreasing trend with $L$, reaching a saturation value that we identify as the bulk gap in the thermodynamic limit (Fig. \ref{Fig4}\textcolor{Navy}{d}).
This length-dependent evolution of the LBE is perfectly in line with the calculated spin spectral weight (purple color map), and can be related to the dispersive nature of triplons. Shorter chains impose a larger value of the minimum wave vector, $k_{\rm min}=\frac{2\pi}{D}$ (with $D=L/2$ for even $L$), pushing the excitation energy up, as illustrated in Fig. \ref{Fig4}\textcolor{Navy}{b}.  
Our measurements for the asymptotic value of the bulk gap, in the limit of long chains ($L>20$), are in perfect agreement with both the triplon-wave expansion theory, equation (\ref{eq3}), and the multiprecision theory methods by Barnes, Riera, and Tennant\textsuperscript{\cite{barnes1999s}}: 
\begin{equation}
	\Delta (\alpha) \approx J_2 (1-\alpha)^{3/4} (1+\alpha)^{1/4}
\end{equation} 
both yielding a bulk gap of $\sim$21.3 meV. 
This quantitative agreement between experiment and theory prediction is particularly noteworthy considering that our spin chain is physisorbed on a metallic substrate (which theory neglects), underscoring the predominance of the exchange interaction over substrate influences.

Finally, we show reversible spin chain control by combining hydrogenative passivation and dehydrogenative activation, which provides a promising avenue for manipulating spin-based quantum devices. Firstly, atomic hydrogen is dosed into the chamber to passivate all spin sites along the chain by the formation of -CH$_2$ group\textsuperscript{\cite{mishra2020topological,Tip-induced}}. Thereafter, tip-induced dehydrogenation is used to remove one of the two hydrogens from the -CH$_2$ group,  which re-activates the spin sites in a controlled way (see \textcolor{Navy}{Methods} for details).
Fig. \ref{Fig4}\textcolor{Navy}{e} illustrates the one-by-one activation for a hydrogenated chain, progressing from $J_1$-terminations ($L=14$) to mixed terminations ($L=15$), $J_2$-terminations ($L=16$), mixed terminations ($L=17$), and finally $J_1$-terminations ($L=18$). 
The corresponding bond-resolved current images confirm the emergent edge spins at $J_1$-terminations, indicated by the drastically increased current owing to the Kondo resonance tunneling (right panel of Fig. \ref{Fig4}\textcolor{Navy}{e}). 
A detailed $dI/dV$ measurement is conducted for the activated segments for each step,  with the ZBC extracted in Fig. \ref{Fig4}\textcolor{Navy}{f} (raw data in Extended Data Fig. \ref{FigE6}). 
The behavior of the edge spin and its oscillating decay in the activated segments mirrors that of the isolated chain, as studied in Fig. \ref{Fig2}\textcolor{Navy}{f}. The gapped feature of the $J_2$ terminations is also evident from the ZBC, with its intensity similar to the bulk. 

Our results highlight the potential for realizing spinful many-body systems in graphene nanomaterials, with energy scales given by chemically tunable exchange interactions. The precise and flexible control over spin sites and interactions may open ways toward the operation of carbon-based quantum spin devices.

\newpage
\section{Methods} 

\begin{figure*}[h]
	\renewcommand{\thefigure}{E1}
	\includegraphics[width=16cm]{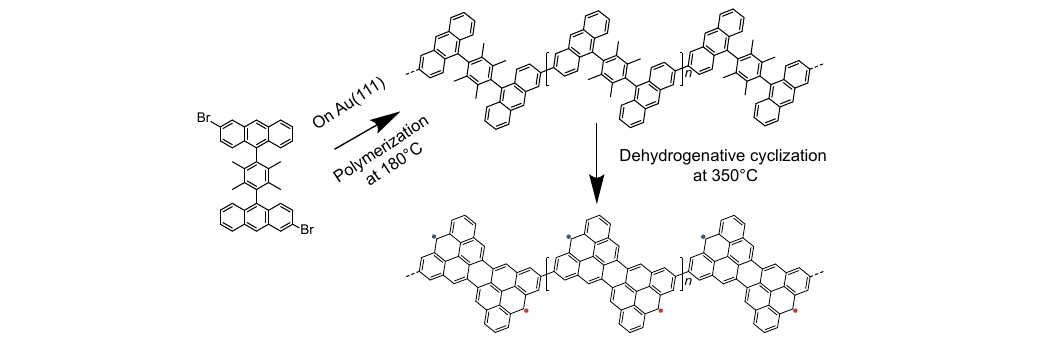}
	\caption{\label{FigE7}
		Schematic drawing for the on-surface synthesis of goblet chains from the precursor to the ribbon.}
\end{figure*}

\begin{figure*}
	\renewcommand{\thefigure}{E2}
	\includegraphics[width=16cm]{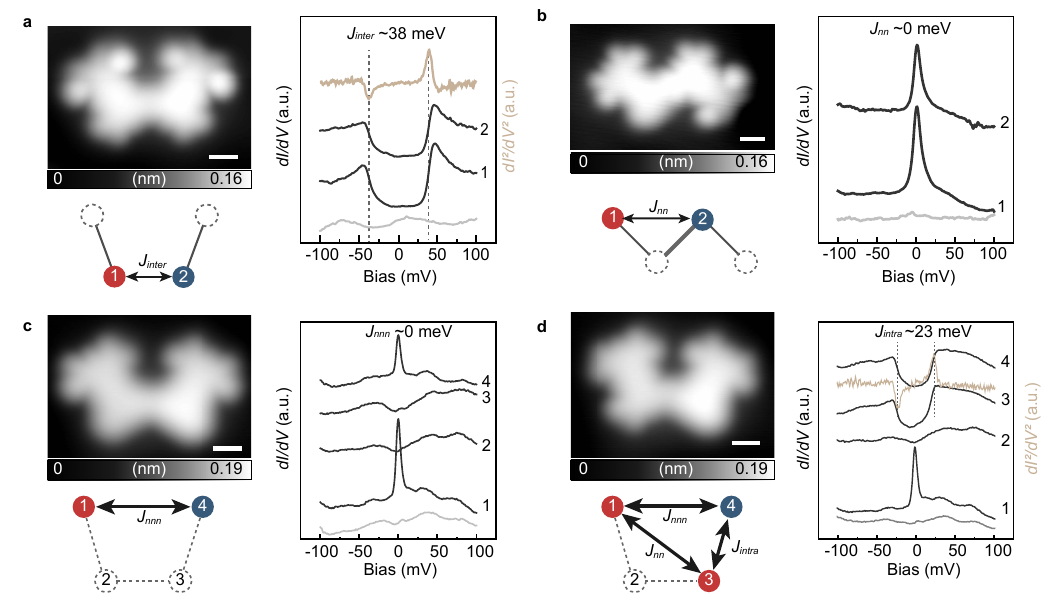}
	\caption{\label{FigE1}
		\textbf{Exchange couplings involved in the goblet chain system}.  
        \textbf{a}, Topographic image of a modified mirror-symmetric dimer containing only the nearest inter-goblet coupling $J_{inter}$. $V_{bias}=0.10$ V, $I_{set}=500$ pA.  The $dI/dV$ spectra taken at different spin sites are displayed in the right panel ($I_{set}=500$ pA), with the averaged $d^2I/dV^2$ spectrum represented as a light brown curve.
        \textbf{b}, Topographic image of a modified dimer containing only the second nearest inter-goblet coupling $J_{nn}$. $V_{bias}=-0.05$ V, $I_{set}=500$ pA.  The $dI/dV$ spectra taken at different spin sites are displayed in the right panel ($I_{set}=500$ pA).
        \textbf{c}, Topographic image of a modified dimer containing only the third nearest inter-goblet coupling $J_{nnn}$. $V_{bias}=-0.05$ V, $I_{set}=500$ pA.  The $dI/dV$ spectra taken at various spin sites are displayed in the right panel ($I_{set}=800$ pA).
        \textbf{d}, Topographic image of an engineered dimer containing $J_{intra}$, $J_{nn}$, and $J_{nnn}$. $V_{bias}=-0.05$ V, $I_{set}=300$ pA.  The corresponding $dI/dV$ spectra are also shown in the right panel, with the averaged $d^2I/dV^2$ spectrum of 3 and 4 represented as a light brown curve. All white scale bars denote 0.5 nm. The background spectra taken on the Au(111) substrate are shown as gray curves. All $dI/dV$ spectra are taken with $V_{rms}=2$ mV.}
\end{figure*}

\begin{figure*}[h]
	\renewcommand{\thefigure}{E3}
	\includegraphics[width=16cm]{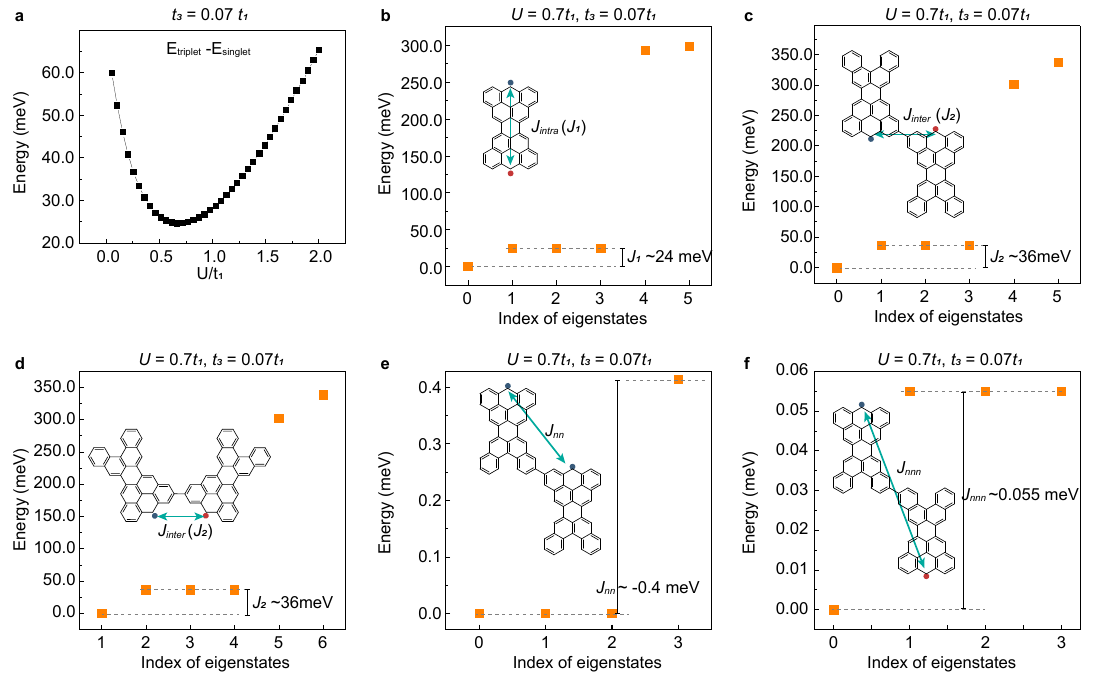}
	\caption{\label{FigE2}
        \textbf{CAS calculation for determining intra/inter-goblet exchange couplings.} \textbf{a}, Excitation energy from the singlet ground state to the triplet excited states of a goblet monomer, as a function of $U/{t_1}$, for $t_3=0.07t_1$, where $U$ is the on-site Coulomb repulsion, $t_1=2.7$ eV is the hopping parameter for the nearest C sites and $t_3$ is the third-neighbor hopping. The parameters that fit the experimental results are found to be $U/{t_1}$ $\sim$ 0.7. \textbf{b}, Eigenvalues of the ground and the first several excited states for a goblet monomer using the optimized parameters.  \textbf{c}-\textbf{f}, Eigenvalues of the ground and the first several excited states for modified goblet dimers having only $J_2$ coupling (inversion symmetric), $J_2$ coupling (mirror symmetric), $J_{nn}$ coupling, and $J_{nnn}$ coupling, respectively.}
\end{figure*}

\begin{figure*}[h]
	\renewcommand{\thefigure}{E4}
	\includegraphics[width=16cm]{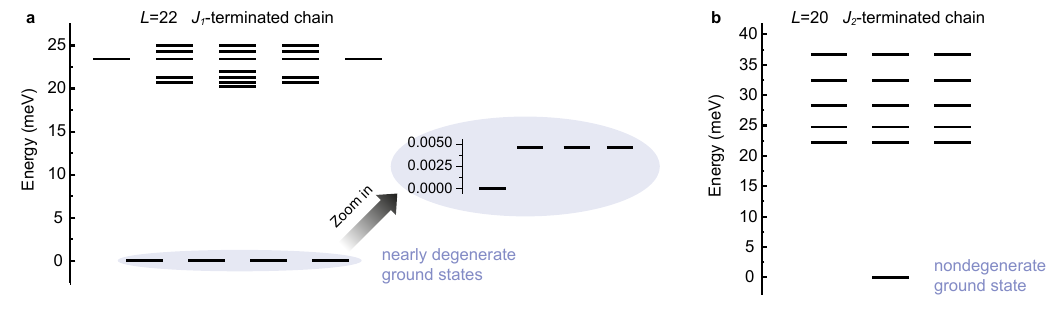}
	\caption{\label{FigE3}
	Calculated eigenstate energies by exact diagonalization (using the QuSpin package\textsuperscript{\cite{weinberg_quspin_2017}}) of chains shown in Fig. \ref{Fig2} in the main text. \textbf{a}, $J_1$-terminated chain with $L=22$. \textbf{b}, $J_2$-terminated chain with $L=20$.}
\end{figure*}

\begin{figure*}[h]
	\renewcommand{\thefigure}{E5}
	\includegraphics[width=16cm]{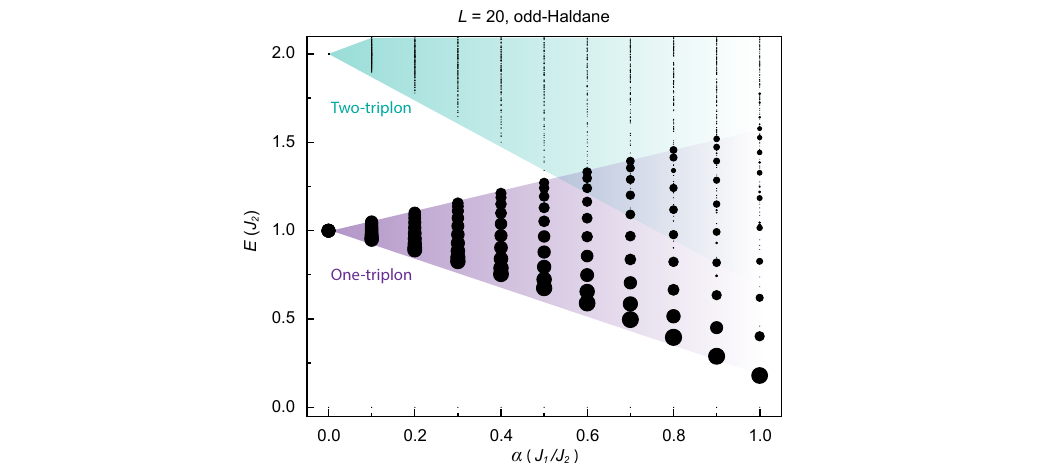}
	\caption{\label{FigE4}
	Calculated excitation energies and probabilities from the ground state to various triplon modes of an $L=20$ odd-Haldane chain through single-site excitation. Different $\alpha$ values are considered ($\alpha=J_1/J_2$). The probability is obtained by summing up the corresponding spin spectral weight of all sites along the chain and is represented by the size of the marker. The one-triplon branch is roughly symmetric around $J_2$ and the bandwidth increases with $J_1$, suggesting that a non-zero $J_1$ is the origin of the dispersive triplon. }
\end{figure*}

\textbf{Sample preparation} Precursor molecules of Clar's goblet (see Extended Data Fig. \ref{FigE7} for detailed structure) were deposited on the clean Au(111) surface via molecular beam epitaxy. Au(111) single-crystal surfaces were prepared by Ar-ion sputtering followed by annealing at 430 $^{\circ}$C for 20 min. During precursor deposition, the molecule source was kept at a temperature of $\sim$ 250 $^\circ$C, and the Au substrate was kept at room temperature. After deposition, we annealed the sample first at $\sim$ 180 $^\circ$C for 10 min for the surface-assisted polymerization and then at $\sim$ 350 $^\circ$C for 5 min for cyclization (see Extended Data Fig.  \ref{FigE7}). Detailed synthetic procedures for the precursor molecule in solution are shown in \textcolor{Navy}{Supplementary Note 4}. 

\textbf{STM and $dI/dV$ measurements} STM and scanning tunneling spectroscopy (STS) $dI/dV$ measurements were performed in a commercial low-temperature STM from Scienta Omicron operating at a temperature of 4.5 K and a base pressure below 2$\times$ 10$^{-11}$ mbar. All the $dI/dV$ spectra were taken with a modulation voltage of 681 Hz by using a CO-functionalized W tip except for the data shown in Fig. \ref{fig1}, where a pure W tip was used. $d^2I/dV^2$ spectra were obtained by numerical differentiation of the corresponding $dI/dV$ spectra. The bond-resolved current images were taken with a CO-functionalized tip in constant-height mode. \textit{In-situ} cold deposition of CO molecules was performed to provide surface-adsorbed CO for tip functionalization.

\textbf{Spin manipulation methods} Both hydrogenation\textsuperscript{\cite{mishra2020topological}} and tip-induced dehydrogenation\textsuperscript{\cite{Tip-induced}} were used for manipulating the spin sites. Hydrogenation is achieved by dosing atomic H with a pressure of $\sim$5$\times$ 10$^{-8}$ mbar into the chamber for 15 min. The atomic H is produced by a thermal gas cracker using high-purity (99.999\%) H$_2$  gas, with working parameters: high-voltage 2000 V, filament current $\sim$4.0 A, electron-beam power $\sim$70 W. The atomic H bonds with specific carbon sites (usually the middle site of the zigzag edges of the goblet which holds the largest spin density) to remove the unpaired electrons by forming sp$^3$ hybridization.  This kind of passivation process is referred to as hydrogenative passivation. The additional H atom (having formed the sp$^3$ configuration)  can be removed by applying a high bias voltage of $\sim$2.0 V using an STM tip. This switches the spin site on again, which is referred to as dehydrogenative activation.  Without previous hydrogenation, tip-induced dehydrogenation can also be used to passivate the spin site by removing the last H on the carbon site, after which the carbon bonds with an Au atom from the substrate and concomitant hybridization and charge transfer effectively remove the unpaired $\pi$-electron.  This is referred to as dehydrogenative passivation.  A schematic illustration for these single-site spin engineering methods is shown in Extended Data Fig. \ref{FigE8}. We compared the isolated $J_2$ values acquired by the hydrogenative passivation and dehydrogenative passivation. In both cases, $J_2$ $\sim$ 38 meV, which confirms the equivalence of these two passivation methods in the framework of the spin model (see Extended Data Fig. \ref{FigE8}).  

\textbf{CAS calculations}
The starting point for the CAS calculations is a single-orbital tight-binding model where we only consider the $p_z$ orbitals of the C atoms that compose the nanographenes.
In our approximation, we consider first and third nearest neighbor hoppings, denoted by $t_1$ and $t_3$, respectively. The corresponding tight-binding Hamiltonian reads as
\begin{align}
    H_{0} = - t_1 \sum_\sigma \sum_{\langle i,j \rangle}  c^\dagger_{i, \sigma} c_{j, \sigma} - t_3 \sum_\sigma \sum_{\langle\langle\langle i,j \rangle\rangle\rangle}  c^\dagger_{i, \sigma} c_{j, \sigma},
\end{align}
where $c^{(\dagger)}_{i,\sigma}$ denotes the annihilation (creation) operator of an electron with spin projection $\sigma=\uparrow,\downarrow$ at site $i$.
This single-particle model is diagonalized, leading to a set of molecular orbitals.
At charge neutrality, a chain with $N$ goblets (i.e., $L=2N$ spin-½ sites) features $2N$ half-filled zero-energy states, slightly hybridized due to $t_3$.
For the CAS calculations, we consider a subset of molecular orbitals that includes these $2N$ zero-energy states, plus the two closest states in energy, necessary to account for the Coulomb-driven exchange mechanism\textsuperscript{\cite{jacob22}}.
Thus, the active space includes $N_{MO}=2N+2$ molecular orbitals.
For clarity, we note that in modified goblet chains we follow a similar approach, where we take into account all the quasi-zero-energy states, plus the two closest states in energy.
Then, we include interactions, within the Hubbard model approximation, where we consider an on-site Hubbard repulsion $U$, given by
\begin{equation}
    H_{U} = U \sum_i n_{i,\uparrow} n_{i,\downarrow},
\end{equation} 
with $n_{i,\sigma} = c^\dagger_{i,\sigma} c_{i,\sigma}$.
The many-body Hubbard Hamiltonian $H_0 + H_U$ is represented in the restricted basis set, considering all the multi-electronic configurations that can be obtained with $N_e$ electrons in the $N_{MO}$ molecular orbitals.
Assuming half-filling, we always have $N_e=N_{MO}$.
The remaining electrons are thus assumed to fully occupy the molecular orbitals below the active space, implying that these are frozen doubly-occupied orbitals; the occupation of the molecular orbitals above the active space is also assumed to be frozen, featuring zero electrons.
Finally, the resulting (truncated) Hubbard Hamiltonian is diagonalized numerically.


\textbf{Modelling $dI/dV$ spectroscopy}
To model the $dI/dV$ spectroscopy we consider a spin chain that is coupled to two electron reservoirs: the STM tip and the substrate. In our description, we consider two types of electron scattering that can produce a spin flip in a given site of the chain\textsuperscript{\cite{Appelbaum1967Exchange}}: electrons that tunnel from tip to sample, exciting the spin chain in the process; and scattering between the substrate electrons. To compute the current, $I$, we apply scattering theory including corrections up to third order\textsuperscript{\cite{ternes2015spin}}; then we differentiate the current with respect to the bias $V$ (defined as the difference between the chemical potential of the two reservoirs), thus obtaining the theoretical prediction for $dI/dV$. Up to second order, the $dI/dV$ spectrum is composed of thermally broadened excitation steps, whose height is determined by the spin spectral weight\textsuperscript{\cite{Rossier2009ITS}}. Excitations are only possible if the total spin, $S$, of the involved states respects the relation $\Delta S=0,\pm1$. The third-order correction accounts for processes mediated by intermediate states; the spin conservation rule is enforced between the initial and final states, as well as between the initial/final state and the intermediate states; energy conservation, however, is only required between the initial and final states. The third order correction is responsible for the introduction of logarithmic resonances which lead to two main changes in the spectrum: i) the thermally broadened steps acquire an overshooting feature at the onset of excitation; ii) a Kondo peak appears at zero bias if the system has a degenerate ground state. In our calculations, we have considered a temperature of 4.5 K.

\textbf{Acknowledgements}
We thank Shantanu Mishra for critical reading of the manuscript.
This work was supported by the Swiss National Science Foundation (grants no. 200020-182015, CRSII5$\_$205987, M.J.$/$PP00P2$\_$170534 and PP00P2$\_$198900), the NCCR MARVEL funded by the Swiss National Science Foundation (grant no. 51NF40-182892), the EU Horizon 2020 research and innovation program-Marie Skłodowska-Curie grant no. 813036, Graphene Flagship Core 3 (grant no. 881603), ERC Consolidator grant (T2DCP, grant no. 819698), the  Center for  Advancing  Electronics  Dresden  (cfaed),  H2020-EU.1.2.2.-FET  Proactive  Grant  (LIGHT-CAP,  101017821), the   DFG-SNSF   Joint   Switzerland-German   Research   Project (EnhanTopo, No. 429265950),  ERC Starting grant (M.J.$/$INSPIRAL, grant no. 716139), the European Union (Grant FUNLAYERS-101079184), Funda\c{c}\~{a}o para a Ci\^{e}ncia e a Tecnologia (Grant No. PTDC/FIS-MAC/2045/2021), Generalitat Valenciana  (Grants No. Prometeo2021/017 and No. MFA/2022/045), MICIN-Spain (Grant No. PID2019-109539GB-C41) and the Advanced Materials programme  supported by MCIN with funding from European Union NextGenerationEU (PRTR-C17.I1). We also greatly appreciate financial support from the Werner Siemens Foundation (CarboQuant). For the purpose of Open Access (which is required by our funding agencies), the authors have applied a CC BY public copyright license to any Author Accepted Manuscript version arising from this submission.

\textbf{Author contributions}
X.F., P.R., and R.F. conceived the project. J.Z. and L.Y. synthesized and characterized the precursor molecules in solution, with supervision provided by J.M. C.Z. and L.Y. performed the on-surface synthesis.  C.Z. performed the STM/STS measurements and analyzed the data. G.C., J.C.G.H., and J.F.R. performed the CAS, ED, and DMRG calculations. C.Z., G.C., J.F.R., and R.F. wrote the Manuscript.
All authors discussed the results.

\clearpage
\bibliography{Goblet}

\newpage
\centering

\section{Extended Data}

\FloatBarrier

\begin{figure*}[h]
	\renewcommand{\thefigure}{E6}
	\includegraphics[width=16cm]{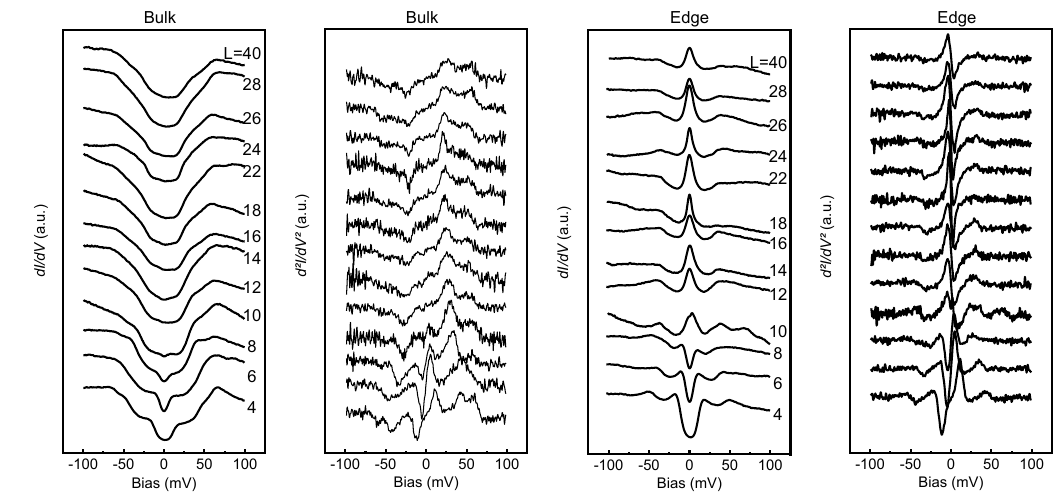}
	\caption{\label{FigE5}
	Bulk (average of spin sites No. $L/2$ and No. $L/2+1$ in the middle) and edge (average of spin sites No. $1$ and No. $L$ at the ends) $dI/dV$ spectra, for chains with different lengths $L$.  For all the $dI/dV$ spectra: $I_{set}=500$ pA, $V_{rms}=2$ mV. The corresponding $d^2I/dV^2$ are also shown aside. Detailed  $dI/dV$ spectra for the long chains can be found in \textcolor{Navy}{Supplementary Note 3}.	}
\end{figure*}

\begin{figure*}[h]
	\renewcommand{\thefigure}{E7}
	\includegraphics[width=16cm]{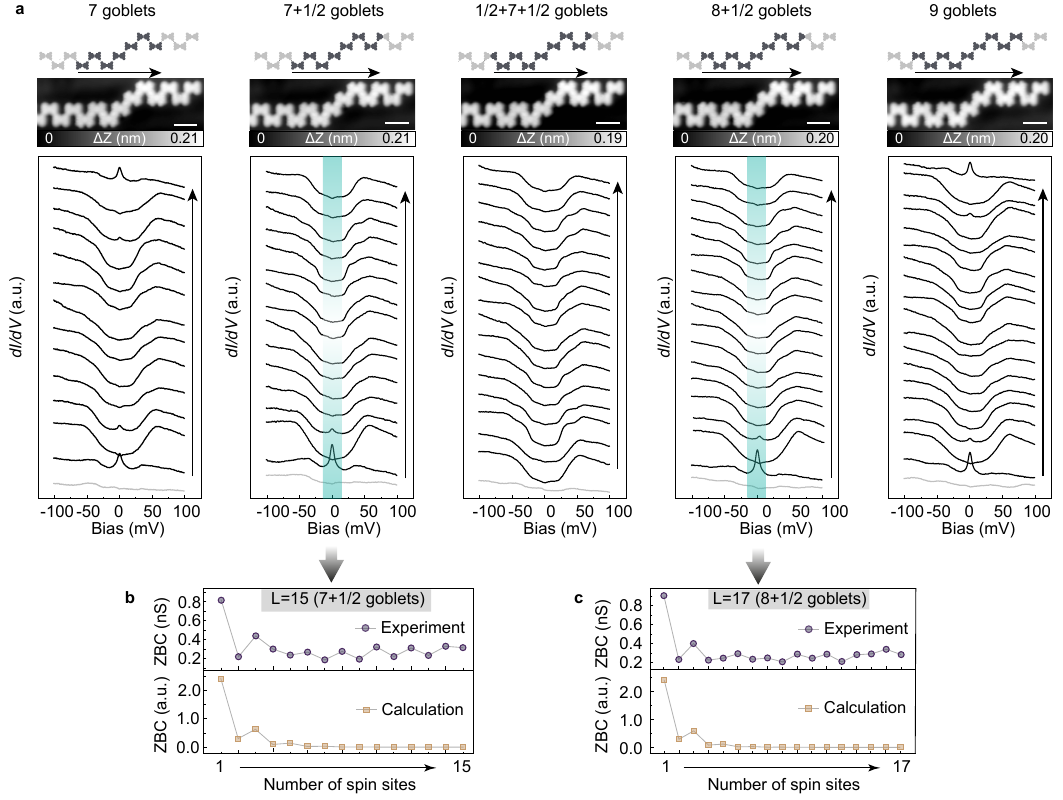}
	\caption{\label{FigE6}
	\textbf{a}, Schematic illustration, STM topographic image, and $dI/dV$ spectra taken on activated spin sites during the step-by-step activation described in Fig. \ref{Fig4}\textcolor{Navy}{e} of the main text. The background spectra taken on the Au substrate are shown by gray curves. All STM images are taken with a CO-functionalized tip: $V_{bias}=-0.1$ V, $I_{set}=100$ pA. All spectra are taken with: $I_{set}=500$ pA, $V_{rms}=2$ mV. \textbf{b} and \textbf{c}, ZBC extracted from $7+1/2$ and $8+1/2$ chains, which have mixed terminations.}
\end{figure*}


\begin{figure*}[h]
	\renewcommand{\thefigure}{E8}
	\includegraphics[width=16cm]{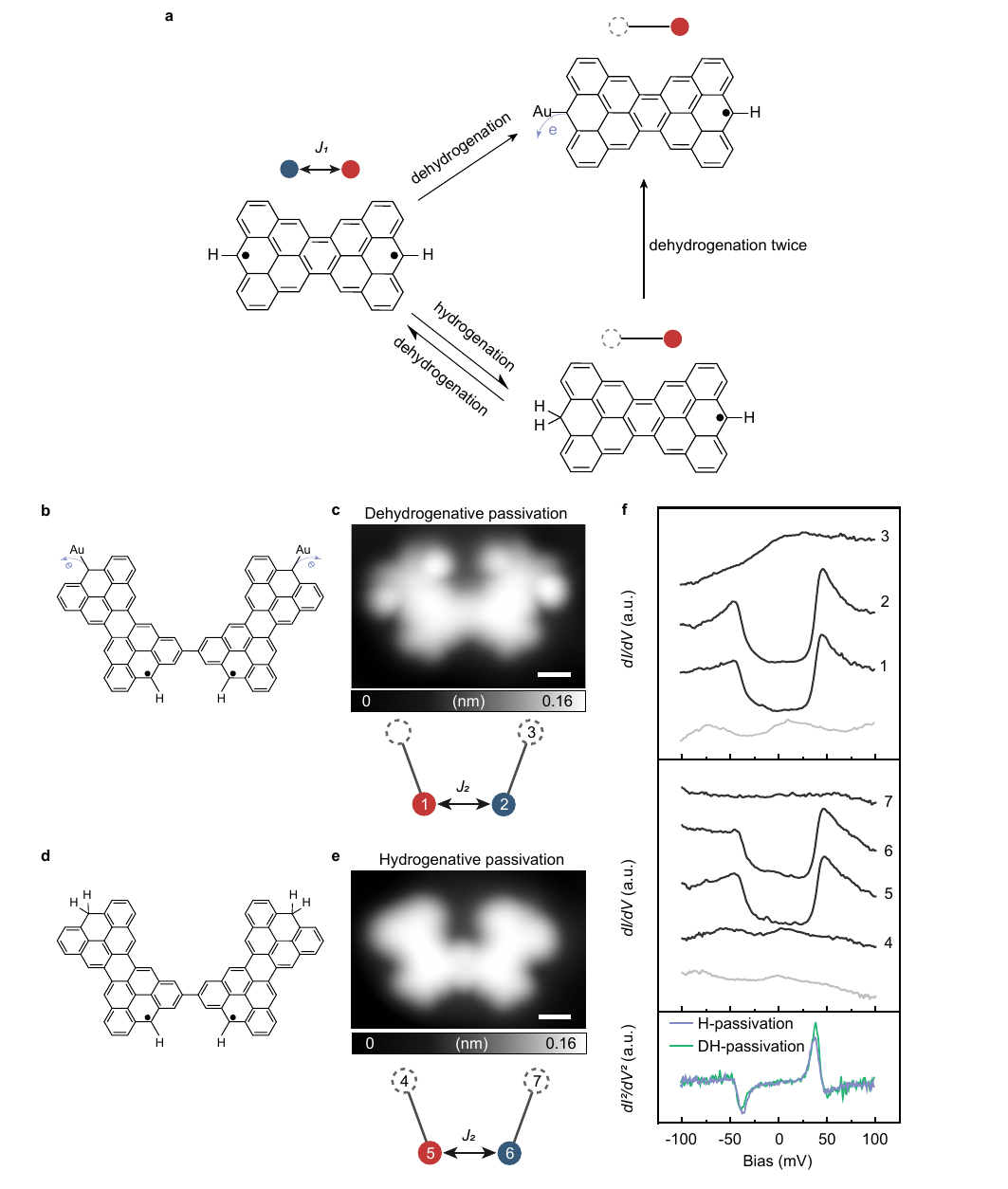}
	\caption{\label{FigE8}
	\textbf{Comparison between dehydrogenative and hydrogenative passivation of spin sites.} \textbf{a}, Schematic illustration of different engineering methods for spin manipulation. \textbf{b} and \textbf{c}, Schematic illustration and topographic image of a dehydrogenative passivated goblet dimer containing only $J_2$. $V_{bias}=0.10$ V, $I_{set}=500$ pA.   \textbf{d} and \textbf{e}, Schematic illustration and topographic image of a hydrogenative passivated goblet dimer containing only $J_2$. $V_{bias}=-45$ mV, $I_{set}=500$ pA.  \textbf{f}, The corresponding $dI/dV$ spectra taken in \textbf{c} and \textbf{e}. $I_{set}=800$ pA. The averaged $d^2I/dV^2$ spectra of 1 (5) and 2 (6) are shown in the bottom panel. All $dI/dV$ spectra are taken with $V_{rms}=2$ mV. Scale bars denote 0.5 nm.}
\end{figure*}



\end{document}